\documentclass[useAMS,usenatbib]{mn2e}

\usepackage[pdftex]{graphicx}

\newcommand{\ch}{${R_h}$}
\newcommand{\chct}{${R_h = c t}$}
\newcommand{\cht}{${R_h = t}$}
\newcommand{\cdm}{${\rm \Lambda CDM}$}

\newcommand{\omegaw}{w}

\title[The Unphysical Properties of the \chct\ Universe]{Matter Matters: Unphysical Properties of the \chct\ Universe}
\author[Geraint F. Lewis]{Geraint F. Lewis\thanks{E-mail:
geraint.lewis@sydney.edu.au}\\
Sydney Institute for Astronomy, School of Physics, A28, The University of Sydney, NSW 2006, Australia
}
\begin{document}

\date{\today}

\pagerange{\pageref{firstpage}--\pageref{lastpage}} \pubyear{2002}

\maketitle

\label{firstpage}

\begin{abstract}
It is generally agreed that there is matter in the universe, and
in this paper, we show that the existence of matter is extremely problematic
for the proposed \chct\ universe. Considering a dark energy component 
with an equation of state of $\omegaw = -\frac{1}{3}$, it is shown that the 
presence of matter destroys the strict expansion
properties that define the evolution of \chct\ cosmologies, distorting the observational
properties that are touted as its success. We further examine whether an 
evolving dark energy component can save this form of cosmological expansion 
in the presence of matter by resulting in an expansion consistent with a mean
value of $\left< \omegaw \right> = -\frac{1}{3}$, finding that the presence of
mass requires  unphysical forms of the dark energy
component in the early universe. We conclude that  matter in the universe
significantly limits the fundamental properties of the \chct\ cosmology, 
and that novel, and unphysical, evolution of the matter component 
would be required to save it. Given this, \chct\ cosmology is not simpler 
or more accurate description of the universe than prevailing cosmological 
models, and its presentation to date possesses significant flaws.
\end{abstract}


\section{Introduction}\label{introduction}
While the \cdm\ picture is the favoured cosmological model, others have suggested
that coincidences in the universe indicate that this picture is incorrect. Instead, it is claimed, 
the universe is in fact a simpler place, with a simple, linear evolution of the expansion of the 
cosmos. Referred to  as the \chct\ universe, it is further claimed that  within this picture, 
key observational features of the cosmos are more accurately explained.

In this paper, we demonstrate that the consideration of the matter content of the universe 
renders the central claims of the \chct\ as invalid, showing that its claimed benefits
over \cdm\ are illusionary, or at least very problematic.

The structure of this paper is as follows:
Section~\ref{cosmichorizon} will briefly present the ideas behind the \chct\ universe, 
whereas Section~\ref{matter}
presents the influence of matter on the evolution of the cosmos. In Section~\ref{observables} 
we discuss the observational consequences of the presence of matter, while in 
Section~\ref{evolving} we ask the question of whether the \chct\ universe can be saved 
by considering an evolving dark energy component. Our conclusions are presented in 
Section~\ref{conclusions}. Throughout we will assume that ${\rm c=1}$, and hence refer to the 
\chct\ cosmology simply as \cht.

\section{The \cht\ Universe}\label{cosmichorizon}
The notion of the \cht\ universe has grown in a series of papers over recent years. 
The foundation
for this cosmological model is an apparent coincidence between the current age of the universe, 
$t_o$, and the current size of the Hubble Sphere, $R_h = \frac{1}{H_o}$, where 
$H_o$ is the present-day Hubble Constant, such that
\begin{equation}
t_o \sim \frac{1}{H_o}
\label{coincidence}
\end{equation}
This coincidence will not generally hold\footnote{As an example, 
recall that in an Einstein-de Sitter 
universe, the age of the universe is related to the  Hubble Constant through $t = \frac{2}{3 H}$, so
an equality in Equation~\ref{coincidence} never occurs.}.
In fact, within the \cdm\ cosmology, $R_h$ increased rapidly in the early epochs 
of the universe, before slowing as the influence of dark energy increases, before asymptoting 
to a particular value in the distant future; within this cosmology, there is a single moment
where the relationship in Equation~\ref{coincidence} holds, and we appear to be close to
it.

One of the earliest comments on this coincidence was provided by \citet{2007arXiv0708.3414L} who noted 
that this implies that the mean acceleration of the universe must be close to zero, and that 
we should be surprised, if we are living in a \cdm\ cosmology, to find ourselves at this rare epoch
\citep[see][]{2010MNRAS.404.1633V}.
Given this, \citet{2007arXiv0708.3414L} further proposed that we can remove this coincidence if the universe
had a simple linear expansion, such that Equation~\ref{coincidence} becomes an exact 
equality at all times.

Through a series of papers~\citep{2007MNRAS.382.1917M,2009IJMPD..18.1113M,2012MNRAS.421.3356B,2009IJMPD..18.1889M,2012JCAP...09..029M}, 
this idea was refined further, firstly with the claim that
the Hubble Sphere, $R_h$, is in fact an unrecognised, yet fundamental, property of the universe, 
labelled the ``Cosmic Horizon"; this is a surprise, as horizons within the standard cosmological
model have been understood for many decades~\citep[see][]{1956MNRAS.116..662R,1991ApJ...383...60H,1993AmJPh..61..883E}.
The initial claim was that $R_h$ represents an infinite redshift surface,
limiting our view of the universe, although this was shown to be incorrect by \citet{2010MNRAS.404.1633V}.
The debate of the Cosmic Horizon has continued in the literature, but its claimed  properties
as a fundamental horizon
have
been shown to be incorrect \citep{2012MNRAS.423L..26L,2013arXiv1301.0305L}.
More recently,  \citet{2012MNRAS.419.2579M} argue that the coincidence 
given by Equation~\ref{coincidence}  is based upon
a deeper significance relating to the ``Weyl Postulate" and the ``corollary to Birkhoff's theorem"
which demands that Equation~\ref{coincidence} is a strict equality that must hold 
at all times (although these arguments were developed through previous papers). If that 
is the case, then the universe has a linear expansion and the Hubble Constant, $H = t^{-1}$.
This is very unlike the decelerating  and then accelerating expansion experienced in the
\cdm\ cosmology, and hence the constituent energy densities in a \cht\ 
universe must be significantly different to the matter and dark energy underlying 
\cdm.

The cosmological implications of an equality of Equation~\ref{coincidence} were further considered, 
especially with regards to  its impact how we observe the universe~\citep[e.g.][]{2012AJ....144..110M,2012arXiv1206.6527M}. As well as claiming that it 
provides a better fit to cosmological observations, one of the key claims about the \cht\ universe is
an apparent underlying simplicity when compared to \cdm~\citep{2012arXiv1205.2713M}. However, 
some of the claims of efficacy of the \cht\ in explaining cosmological observations have been called into question~\citep{2012MNRAS.425.1664B}. 

This question of the make-up of the \cht\ universe has been examined in several 
papers, most recently by \citet{2012MNRAS.419.2579M}, who concluded that the energy density
driving the expansion must possess a time-averaged equation of state of 
$\left<\omegaw\right> = -\frac{1}{3}$, although if an equality in Equation~\ref{coincidence} is 
required, then the equation of state must be $\omegaw = -\frac{1}{3}$ at all times.

For a universe with a critical density of a single fluid,  an equation of state of 
$\omegaw = -\frac{1}{3}$ represents a dividing line between energy densities 
that decelerate cosmic expansion $(\omegaw > -\frac{1}{3})$, and those that result
in universal acceleration $(\omegaw < -\frac{1}{3})$, and hence results in a 
uniform cosmic expansion.  

The real universe, however, possesses a mix of energy densities; the \cdm\ universe, 
currently possess two dominant components, matter (with $\omegaw = 0$) and 
a cosmological constant ($\omegaw = -1$). Irrespective of the fact that we may not
fully understand the properties of the dark energy component, it is pretty 
uncontroversial that there is matter in the universe, either the substantial quantity 
of dark matter inferred from dynamical arguments, or the more modest baryonic
component we can directly observe. Hence, how is the requirement of $\omegaw = 
-\frac{1}{3}$ to be interpreted in the presence of matter? This is the subject of the remainder of this paper.

In the following, we will adopt the tenet of the \cht\ universe, namely that is it simple when compared
to other cosmologies, such as \cdm~\citep{2012arXiv1205.2713M}. 
We will assume that the expansion of the universe is governed by Einstein's general relativity through 
the Friedmann equations; we note that
there is potential for the universe experiencing \cht\ evolution if gravity acts only locally, without a long-range 
interaction, although this would not be consistent with the standard cosmological model, and is, therefore,
beyond the scope of this paper.
As well as \cdm, we also consider universes that contain 
a matter component and a dark energy component with an equation of
state $\omegaw = -\frac{1}{3}$; at late times, when the density of matter has dropped to 
negligible amounts, these cosmologies will asymptote to  \cht\ universes. 
Later, we will relax this assumption and allow the equation of state
of dark energy to evolve over cosmic history, such that the accelerated 
expansion of the dark energy counterbalances the deceleration due to the
presence of matter, and the result expansion is that demanded by the \cht\
cosmology. We will examine the consequences of this on the resultant 
required form of the dark energy equation of state.

\begin{figure}
\includegraphics[scale=0.45, angle=0]{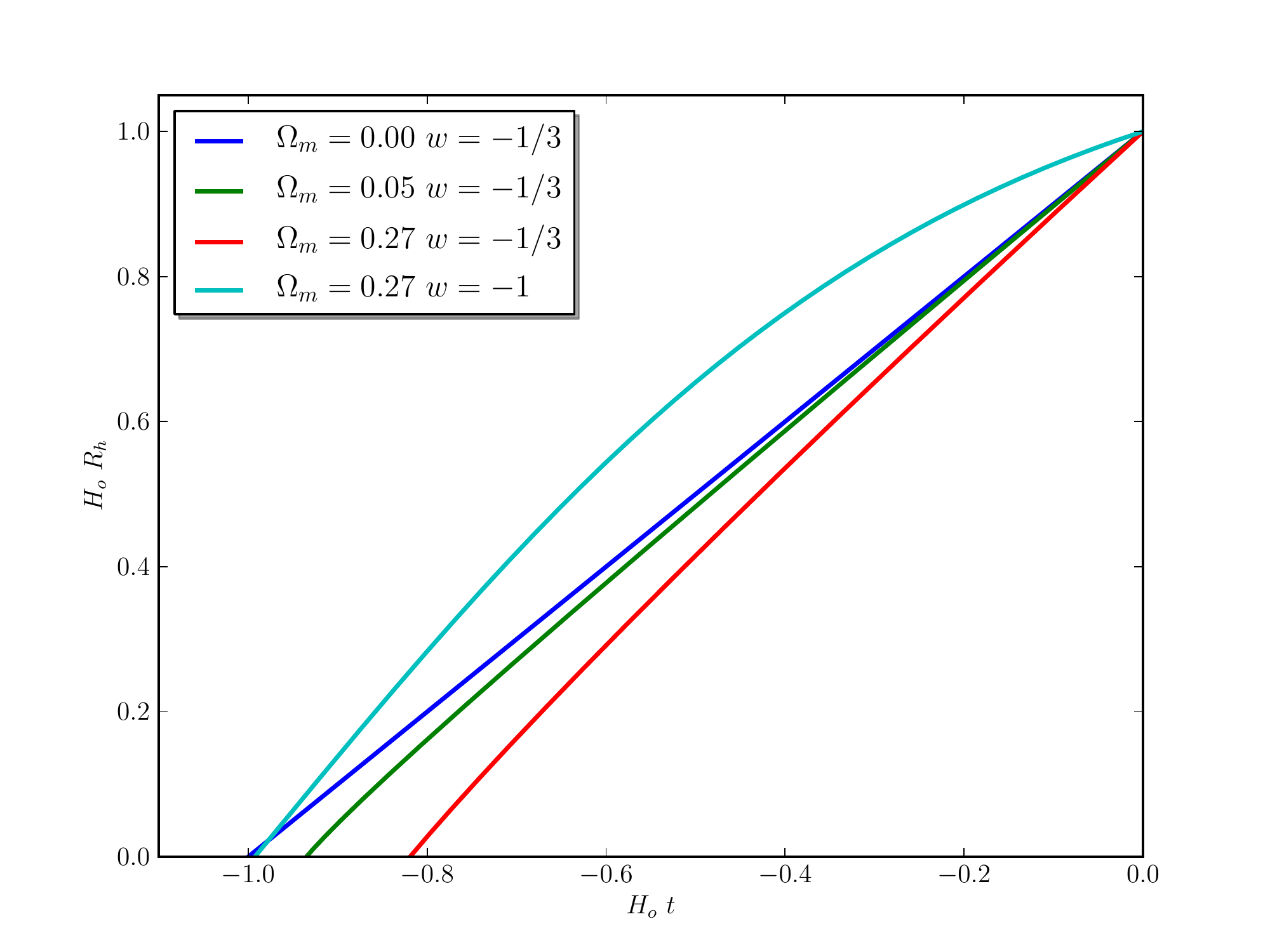}
\caption{The cosmic horizon, \ch, for a range of cosmological models, as a function of cosmic time. 
The dark blue line corresponds to 
a mass-less universe, with only a dark energy component with $\omegaw = -\frac{1}{3}$, whereas the 
green and red lines correspond to similar cosmological models but with present day normalised mass
densities of 0.05 and 0.27 respectively.  The cyan line corresponds to a \cdm\ universe, 
with $\omegaw=-1$.  
\label{figure1}
}
\end{figure}

\section{The Influence of Matter}\label{matter}
To understand the influence of matter on the expansion of the universe, we start with 
the Friedmann-Robertson-Walker (FRW) invariant interval for a spatially flat 
cosmological space-time, given by;
\begin{equation}
ds^2 = -dt^2 + a(t)^2 \left( dr^2 + r^2 d\Omega^2 \right)
\label{frw}
\end{equation}
where $t$ is the cosmic time, $r$ is the radial comoving coordinate and $d\Omega$ accounts for
angular terms. The normalised scale-factor, $a(t)$ evolves with cosmic time and embodies the
evolving cosmos.

In the following, we will consider a universe containing two energy density components, one 
matter, with an equation of state of zero, and a dark energy component with an equation of state of 
$\omegaw$. The normalised present-day densities of these components will be given $\Omega_m$ 
and $\Omega_{de}$ respectively \citep{1988A&A...206..175L}; in the following we will neglect
the influence of the photon energy density which dominates in the earliest epochs of the universe, 
but as this results in a larger deceleration during the time it dominates it  only exacerbates the 
problems outlined in this paper.

The Friedmann equations tell us that the the scale-factor in a spatially-flat universe will evolve according to
\begin{equation}
H \equiv \frac{\dot{a}}{a} = H_o \sqrt{ \Omega_m a^{-3} + \Omega_{de} a^{ -3( 1 + \omegaw ) } }
\label{hubble}
\end{equation}
where $H$ is the Hubble constant, with a present day value of $H_o$.
It is straight-forward to integrate Equation~\ref{hubble} and determine the evolution of $a$ and
then calculate $R_h$ as a function of cosmic time, 
and in Figure~\ref{figure1} we present this for several fiducial 
cosmologies. Note that for this plot, zero on the time axis corresponds to the present day, and 
we integrate backwards to the Big Bang in each model.

The blue curve presents the pure \cht\ universe, with no matter, so $\Omega_m = 0$, and
consisting of only a dark energy component, with $\Omega_{de} = 1$, with an equation of state
of $\omegaw=-\frac{1}{3}$. As expected, this possesses a linear evolution from zero, at a time of
$H_o t = -1$ to a value of unity today. For comparison, we also present the evolution of $R_h$ 
in \cdm\ (in cyan); clearly this does not possess such a linear evolution.
 
\begin{figure}
\includegraphics[scale=0.45, angle=0]{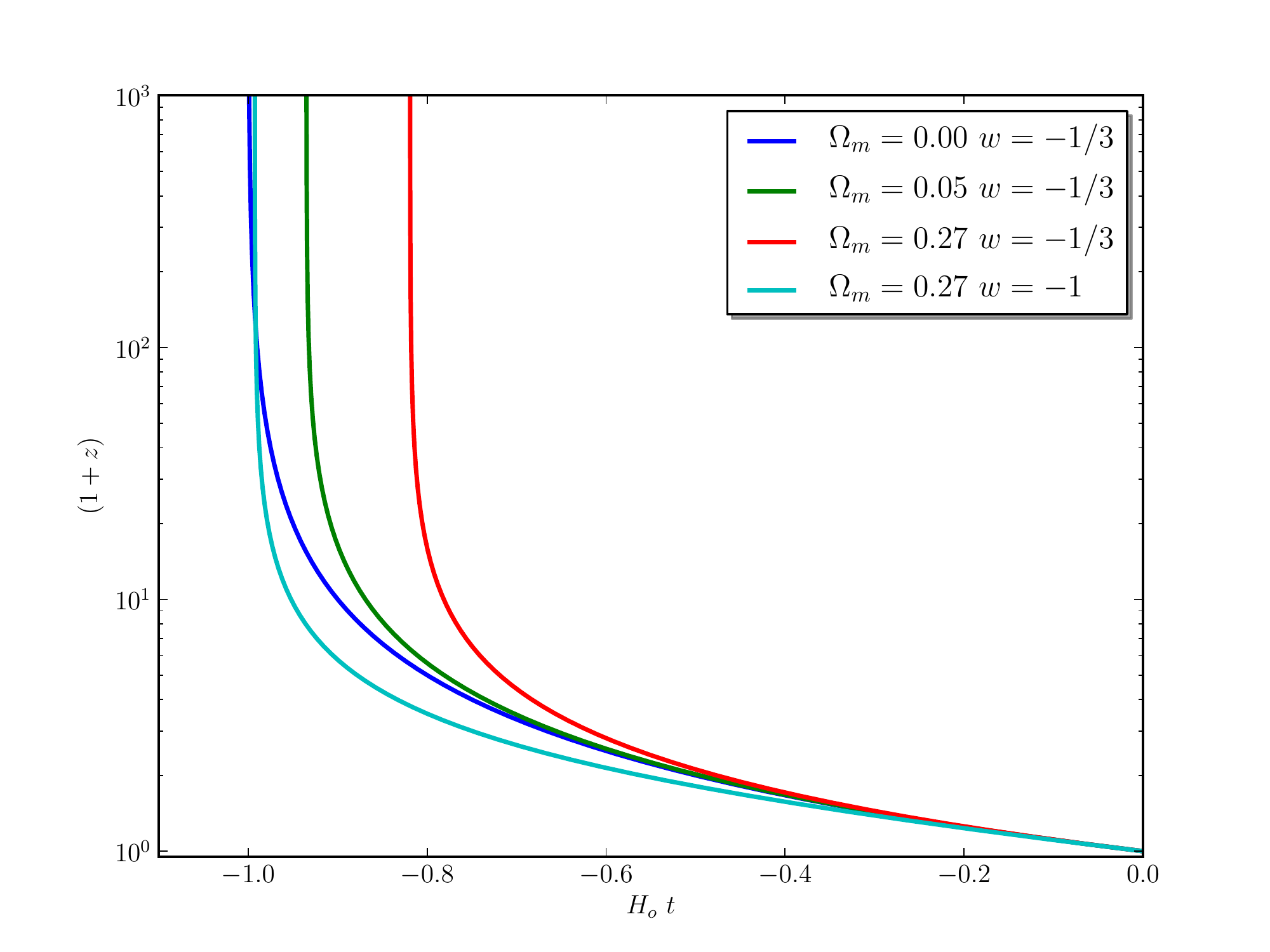}
\caption{As Figure~\ref{figure1}, but presenting the redshift, as observed today, as a function of cosmic time.
\label{figure2}
}
\end{figure}

The remaining two lines replace some of the dark energy in the \cht\ cosmology with matter.
In the first case (red line) we adopt the currently favoured present day matter density of $\Omega_m = 0.27$, 
such that $\Omega_{de} = 0.73$ but assuming an equation of state of $\omegaw = -\frac{1}{3}$,  
rather than $\omegaw = -1$ of \cdm.
While this trend is roughly linear, clearly it is not the required
$R_h = t$ evolution of the Cosmic Horizon that is apparently demanded by the ``Weyl Postulate" 
and the corollary of ``Birkhoff's theorem"~\citep{2012MNRAS.419.2579M}.

With the final green line in Figure~\ref{figure1}, we banish dark matter and assume that the only
matter present is the baryonic component we can observe, adopting an $\Omega_m=0.05$. Again, 
the evolution of the Cosmic Horizon is roughly linear, and, while it approximately coincides with 
\cht\ at  the present time, it clear deviates from this relation in the earlier epochs of cosmic
history.

One conclusion one can immediately make from an examination of Figure~\ref{figure1} is that 
if the dark energy component of the universe possesses an equation of state of $\omegaw = -\frac{1}{3}$
then the presence of {\it any} quantity of matter will result in a deviation from the strictly 
required evolution of the \cht\ universe. While in Section~\ref{evolving} we will consider if an
evolving equation of state of the dark energy component can result in a \cht\ expansion 
history, we will now examine the observational consequence of matter in dark energy universes
with $\omegaw = -\frac{1}{3}$.

\section{Observational Consequences}\label{observables}
This paper will consider only two of the recent claims of the observational superiority of the 
\cht\ universe over other cosmological models. This will be the relation between look-back
time and redshift, and the lack of the requirement on an inflationary period to explain the observed
casual properties of the Cosmic Microwave Background (CMB).

\subsection{Look-back time verses redshift}\label{lookback}
A recent claim of the success of the \cht\ is that it solves the problem of the existence of 
high redshift quasars, given the limited time for the growth of supermassive back holes since the
Big Bang \citep{2013arXiv1301.0017M}. Here we present an identical analysis for the
cosmologies considered in Section~\ref{matter}.

\begin{figure*}
\includegraphics[scale=0.9, angle=0]{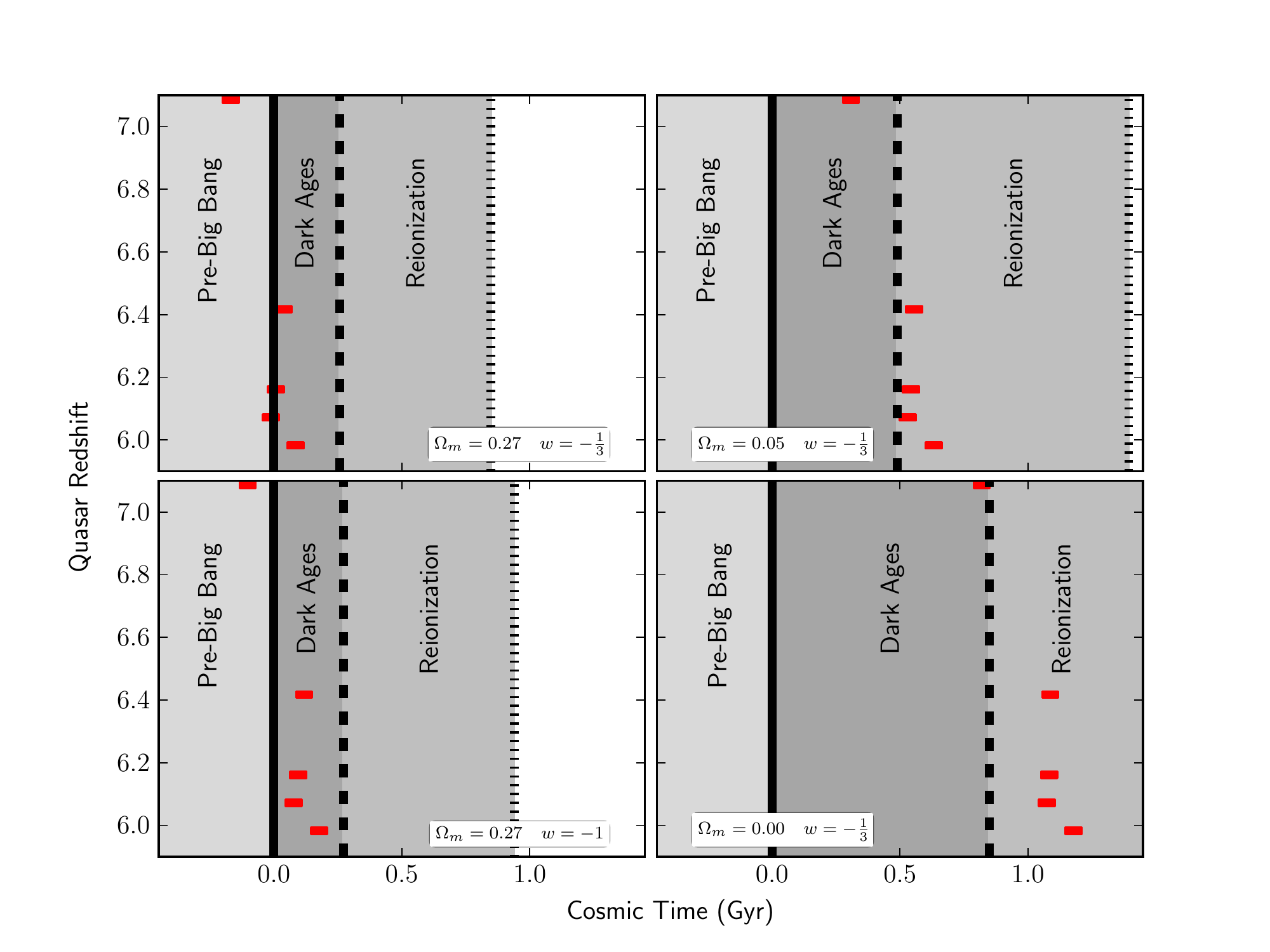}
\caption{
The initial seed period for the formation of super-massive black holes
for the universes considered in this paper,
high-lighting the period of reionization $(z \sim 5-16)$. This presents a subsample of quasars from
\citet{2013arXiv1301.0017M}, 
but uses the same formation time-scales for 5 ${\rm M_\odot}$ and 20 ${\rm M_\odot}$ seed black holes.
The Big Bang is marked as a solid black vertical line, whereas the end of the Dark Ages is given by a 
dashed line. The red horizontal bars are the formation epochs for seed black holes as described 
above. The left-hand lower panel, with $\Omega_m = 0.27$ and $\omegaw = -1$, corresponds to a
\cdm\ universe, whereas the right-hand panel, with $\Omega_m = 0.00$ and $\omegaw = -\frac{1}{3}$
corresponds to the \cht\ universe. The upper panels present cosmologies with a dark energy component
with $\omegaw = - \frac{1}{3}$, but with $\Omega_m = 0.27$ (left-hand panel) and $\Omega_m = 0.05$ 
(right-hand panel).
\label{lookback2}
}
\end{figure*}

The two key ingredients for this analysis are the age of the universe at the epoch of 
quasar activity, given by the look-back time verses redshift given in 
Figure~\ref{figure2}~\footnote{It appears several numerical errors have crept into the analysis of
\citet{2013arXiv1301.0017M}. 
For example, 
for $H_o = 72\ km/s/Mpc$, the age of the universe in the \cht\ cosmology is 13.58 GYrs, and
therefore the age of the universe for ULAS J1120+0641 at a redshift of $z=7.085$ \citep{2011Natur.474..616M} 
should be 13.58 Gyrs /(1+7.085) = 1680 Myrs, not the 1634 Myrs listed in the paper.}, while
the other is the time-scales for the formation of supermassive black holes. As with 
\citet{2013arXiv1301.0017M}, rather than black holes being primordial, 
we will assume that supermassive black holes  grow 
from  seed
black holes of masses of either ${\rm 5 M_\odot}$ or ${\rm 20 M_\odot}$, resulting from supernovae 
in the first generation of stars.
Following \citet{2013arXiv1301.0017M}
we adopt the relation between the seed mass, $M_o$, and the time, $\tau_{M_o}$ for the growth to 
a mass of $M$ to be
\begin{equation}
M = M_o \exp \left( \frac{\tau_{M_o}}{ 45\ {\rm Myrs}} \right)
\label{timescale}
\end{equation}
Hence, given the time that we see a quasar, and an estimate of its black hole mass, it is straight forward
to calculate the time at which appropriate seed mass black holes were required.

Directly taking a subsample of the 
tabulated~\footnote{While it is not clearly described in \citet{2013arXiv1301.0017M}, the 
Seed/Myr columns of their Table 1 corresponds to the ages that you would require appropriate mass
seed black holes in the \cht\ universe. The values for \cdm\ are not given.} 
values from \citet{2013arXiv1301.0017M}, the lower left-hand panel of 
Figure ~\ref{lookback2} illustrates the problem in \cdm; the 
red bar corresponds to the epoch in which seed black holes of masses between ${\rm 20 M_\odot}$ and
${\rm 5 M_\odot}$ need to be formed to result in the quasars we see (this is equivalent to Figure~1 in
\citet{2013arXiv1301.0017M}). Clearly, in a \cdm\ universe,
seed black holes need to be formed in the ``Dark Ages", before the supernovae required to create them.
The situation is even more extreme for the highest redshift quasar, ULAS J1120+0641 at $z=7.085$
\citep{2011Natur.474..616M}, which requires seed black holes to be formed before the Big Bang. 
Various solutions, such as super-Eddington accretion \citep[e.g.][]{2006ApJ...650..669V}, have been proposed
to allow supermassive black holes to grow in the limited time available. 

The lower right-hand panel  in Figure~\ref{lookback2} presents the same situation in the \cht\ universe. With the different 
relationship between cosmic time and redshift (Figure~\ref{figure2}), the seed formation time for quasars now occurs later
in the universe, most occurring well after dark ages, during reionization when massive stars are
readily available to produce black hole seeds.

While this may appear to be a success for the \cht\ universe, it is clear that
from Figure~\ref{figure2} that the look-back time verses redshift relations are significantly
different when matter is added to a $\omegaw = -\frac{1}{3}$ universe. In the
upper right-hand panel of Figure~\ref{lookback2}
we present the case when the present matter density is $\Omega_m = 0.05$, and 
as can be seen, this moves epoch for the production of black hole seed back to just 
after the dark ages which, while better than the \cdm\ universe, reduces some of 
the claimed advances of a pure \cht\ universe.

The situation is exacerbated when considering a more realistic present day matter 
density of $\Omega_m = 0.27$ (upper left-hand panel of Figure~\ref{lookback2}), as the formation time
for the black  hole seeds are now pushed back well into the dark ages, with the
requirement that some seed black holes are required at the epoch of, or even before, the Big Bang. 
Hence, we can conclude that presence of matter significantly curtails the
claimed advantage of the \cht\ universe in providing ample time for the production
of supermassive black holes in the early universe, and that the question of how these
grew in the limited time available is still to be answered.

\subsection{The need for inflation?}\label{inflation}

\begin{figure*}
\includegraphics[scale=0.9, angle=0]{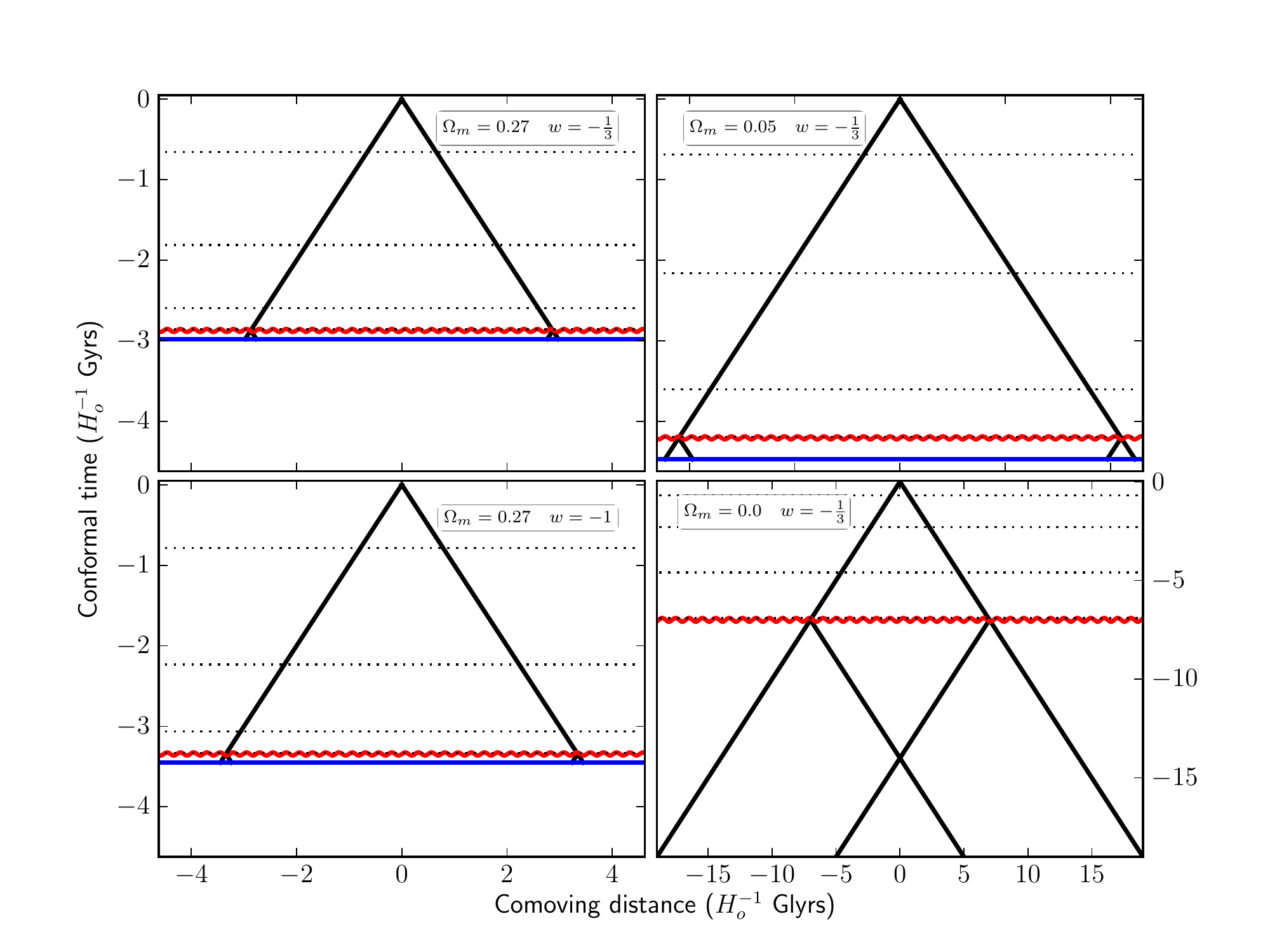}
\caption{The history of the universes considered in this paper in conformal coordinates.
The abscissa presents the comoving distance, while the left-hand ordinate 
presents the conformal time, integrated backwards from the present time $(\tau=0)$.
The large black triangle represent our past light cone, with us at the apex. The blue 
horizontal line represents the Big Bang (a finite conformal horizon; see 
Equation~\ref{conformal}), whereas the red line represents the epoch of recombination.
The dotted lines represent epochs corresponding to $( 1 + z ) = 2, 10, 100\ \&\ 1000$ (from 
top to bottom). As in Figure~\ref{lookback2}, the left-hand lower 
panel, with $\Omega_m = 0.27$ and $\omegaw = -1$, corresponds to a
\cdm\ universe, whereas the right-hand panel, with $\Omega_m = 0.00$ and $\omegaw = -\frac{1}{3}$
corresponds to the \cht\ universe.
The black triangles at the base represent the past light cones for the surface of 
recombination;  for three of the universes, these do not 
overlap before they hit conformal horizon, the 
Big Bang, these points at recombination were never in causal contact, but for the \cht\ 
universe, where the Big Bang occurs at a conformal time of $\tau = -\infty$, these light cones 
overlap and all points were in causal contact before recombination. Note the difference in 
axes scales for the panel presenting the \cht\ universe.
\label{conformalCDM}
}
\end{figure*}

Another grand claim of the \cht\ cosmology is that does away with the need for inflation 
in explaining the causal contact of differing patches during the earliest stages of the universe~\citep{2012arXiv1206.6527M,2012arXiv1207.0734M,2012arXiv1207.0015M}. 
This is more clearly understood in examining the expansion of the universe in a conformal 
representation~\citep[see][for a full explanation]{1991ApJ...383...60H}. Essentially, for 
this representation the abscissa is the comoving radial coordinate, while the ordinate 
is the conformal time, given by
\begin{equation}
\tau = \int_{t_o}^{t_e} \frac{dt}{a(t)}
\label{conformal}
\end{equation}
where $t_o$ is the present epoch and $t_e$ is a different epoch in the universe.
In this coordinate system, light-rays have the same $45^o$ paths as in flat Minkowski space-time, 
making the understanding of causal connexions straight-forward.

The first key property to understand with regards to the need for an inflationary epoch in the universe 
is whether the cosmic evolution results in a finite or infinite `conformal horizon' when integrating 
back to the Big Bang, $t_e = 0$.
If we consider a spatially flat universe consisting of a single fluid with an equation of state, $\omegaw$, then it is
simple to show that
\begin{equation}
a(t) \propto t^\frac{2}{3(1+\omegaw)}
\label{expansion}
\end{equation}
and the integral in Equation~\ref{conformal} in the limit of $t_e \rightarrow 0$ becomes
\begin{equation}
\tau = \kappa \left[ t^{\left( 1 - \frac{2}{3 ( 1 + \omegaw)} \right)} \right]_{t_o}^0
\label{eq}
\end{equation}
where $\kappa$ is a constant and $t_o$ is the present age of the universe.
The zero limit of the integral, corresponding to the Big Bang at $t_e=0$, means that
this integral converges {\it iff} $\omegaw > -\frac{1}{3}$. 
When considering an observer in such a universe, this convergence of the integral 
implies that any past light cone can be extended backwards a finite conformal time before
they reach the Big Bang. However, in universes where $\omegaw \leq -\frac{1}{3}$, the conformal
integral does not converge, and past light cones can be extended backwards an arbitrarily 
large amount of conformal time, encountering the Big Bang at $\tau = -\infty$. Such universes
do not possess particle horizons, with immediate consequences for causality
\citep{1991ApJ...383...60H}~\footnote{The situation is reversed when we consider future conformal
horizons, with Equation~\ref{conformal} converging for $t_e \rightarrow \infty$ for $\omegaw < -\frac{1}{3}$,
and these universes possess an ``event horizon". Hence, the \cht\ universe possesses neither a particle or
event horizon. However, a \cdm\ universe possess both, being matter (and radiation) dominated at early 
epochs, and by dark energy with $\omegaw = -1$ in the future (see panel 3 of Figure 1 in \citet{2004PASA...21...97D}
for a conformal representation of this cosmology).
}

In examining the importance of the existence of a finite conformal horizon, it is useful to look at
the \cdm\ cosmology, as can be seen in the lower left-hand panel of Figure~\ref{conformalCDM} which presents
the conformal structure of a universe with a present day matter density of $\Omega_m=0.27$ and an
equation of state of dark energy given by $\omegaw=-1$. While dark energy is the dominant component 
today, in the past matter dominated, and this universe possesses a finite conformal horizon at the
Big Bang, denoted by the blue line. The red line corresponds to the epoch of recombination, the
source of the Cosmic Microwave Background at a redshift 
of $z\sim 1100$ with respect to a current observer. The large black triangle corresponds to the past light 
cone of an observer today, stretching back to the Big Bang, whereas the smaller black triangles at the 
are the past light cones for emitters at recombination. These smaller light cones do not overlap, and
hence these two points on the CMB were never in causal contact, and have no reason (other than 
fine tuning) to be of the same temperature; this is simply a restatement of the well known Horizon
Problem.

How does this picture change if we consider the \cht\ universe? As mentioned previously, a universe
containing a single fluid at critical density possess a past conformal horizon if $\omegaw > -\frac{1}{3}$, and 
hence, if the universe were solely a fluid with $\omegaw = -\frac{1}{3}$ then all past light cones can become
infinite in extent. Recombination, however, occurs at a finite conformal time in the past, and hence all
light cones from this epoch can be extended backward by an arbitrary amount and all overlap; in such a
model, all points at recombination were in causal contact and hence an inflationary epoch is not required; 
this is represented in the lower right-hand panel of Figure~\ref{conformalCDM}.

However, the situation is different with the addition of matter. In the upper left-hand panel of 
Figure~\ref{conformalCDM}, the matter 
density remains the same as in the \cdm\ universe, but now the equation of state of dark energy
is given by $\omegaw=-\frac{1}{3}$. The overall structure in this cosmology is very similar to that seen in 
the lower right-hand panel of 
Figure~\ref{conformalCDM}, with the existence of a finite conformal horizon in the Big Bang, due
to matter becoming dominant in the early universe, and two relatively 
small light cones connecting this with the epoch of recombination. Again, these past light cones do not 
overlap and hence the points on the surface of recombination were not in causal contact, and so a 
mechanism similar to inflation is required to solve the Horizon Problem.

The situation is not significantly improved by reducing the present day matter density further to
$\Omega_m = 0.05$, as presented in the upper right-hand panel of 
Figure~\ref{conformalCDM}. Again, the structure of the conformal 
representation of the universe is relatively unchanged, with very similar features observed previously, 
and again, the past light cones from recombination back to the Big Bang do not overlap, revealing no
previous causal contact and hence still requiring a solution to the Horizon Problem.

The universal behaviour for the universes with any matter in Figure~\ref{conformalCDM} is 
easy to understand as when we look at earlier and earlier epochs in the universe, matter becomes more and 
more dominant, significantly more than the influence of  the dark energy component. In all of these cases, 
this dominant dark energy component results in a finite conformal horizon when integrating 
Equation~\ref{conformal} back in time, and if there is not sufficient conformal time between the Big Bang 
and recombination, past light cones will not have the opportunity to overlap; this is precisely what 
inflation does \citep[see Figure 2 in][]{1991ApJ...383...60H}.

\section{Evolving Dark Energy?}\label{evolving}
Can the $R_h = t$ universe be saved if we allow the equation of state of the dark energy component 
to differ from $\omegaw=-\frac{1}{3}$? Clearly, $\omegaw$ cannot be a constant over cosmic history, because 
if $\omegaw < 0$, then the universe will always become matter dominated in its earliest epochs, and if 
$\omegaw > 0$, the resulting dominant energy component in the early universe will result in more significant 
deceleration and deviation from $R_h = t$. Hence, in the following, we will search for a evolving 
$\omegaw$ such that the required expansion is achieved.

\begin{figure}
\includegraphics[scale=0.45, angle=0]{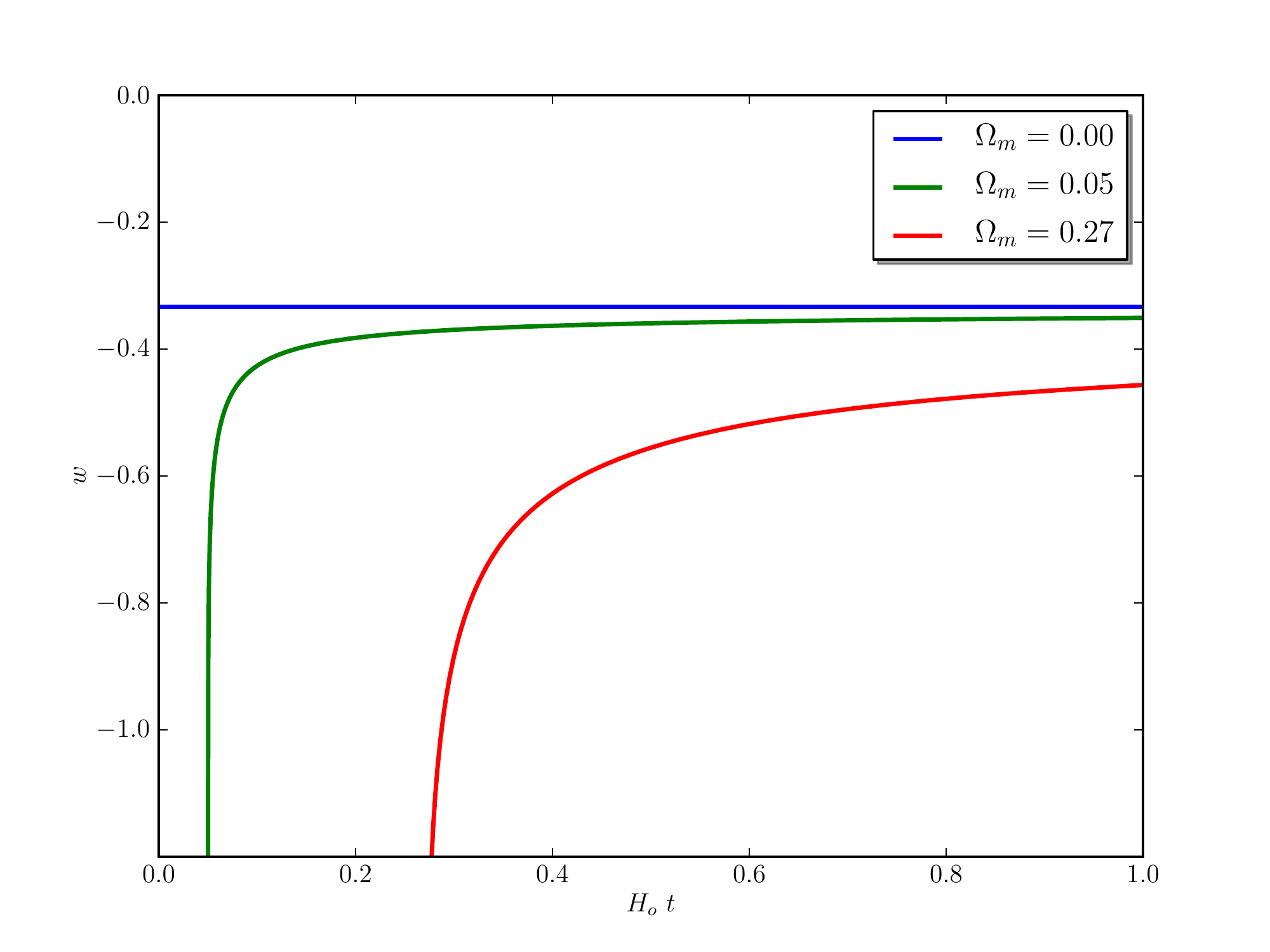}
\caption{The evolving equation of state, $\omegaw$, as a function of cosmic time, required to ensure that $R_h = t$. 
With the presence of any non-zero matter component, $\omegaw$ becomes unphysical at sufficiently early times 
(see Equation~\ref{equality}).
\label{figure3}
}
\end{figure}

Remembering that for $R_h = t$, then $\dot{a} = H_o$ and $a = H_o\ t$ (which is found for a spatially flat 
universe with a single dark energy component with $\omegaw = -\frac{1}{3}$), then
\begin{equation}
a \sqrt{ \Omega_m a^{-3} + \Omega_{de} a^{ -3( 1 + \omegaw ) } } = 1
\label{relation}
\end{equation}
Rearranging this expression, we can solve for $\omegaw$ as a function of cosmic time, $t$, that will ensure
that the expansion results in $R_h = t$ at all times;
\begin{equation}
\omegaw = -\frac{1}{3} \left(  1 + \log{ \left[ \frac{ 1 - \Omega_m / (H_o t ) }{\Omega_{de}} \right]}/ \log{\left[ H_o t \right]}   \right)
\label{w_evolve}
\end{equation}
If $\Omega_m = 0$ and $\Omega_{de} = 1$, then $\omegaw = -\frac{1}{3}$ at all times, but for any $\Omega_m \neq 0$
then $\omegaw$ evolves over cosmic history.
A closer examination of Equation~\ref{w_evolve}, however, reveals that a real solution is only possible {\it iff} 
\begin{equation}
1 - \Omega_m / (H_o t ) > 0
\label{equality}
\end{equation}
such that it is required that $H_o t > \Omega_m$; Figure~\ref{figure3} presents this evolution for several 
fiducial cosmologies. As expected, with no matter content $( \Omega_m = 0 )$, the equation of state of the
dark energy is constant at $\omegaw = -\frac{1}{3}$. However, in the other two cases, with $\Omega_m = 0.05$ 
(green line) and $\Omega_m = 0.27$ (red line), then $\omegaw$ significantly deviates from this value. As
we go back into the past, in models with matter, $\omegaw$ becomes more and more negative, and in 
both models cross the `Phantom Divide' $(\omegaw = -1)$, before diverging to $\omegaw = -\infty$. 
Hence, 
given the inequality in Equation~\ref{equality}, in any universe with any matter content, there is no evolving equation of
state for dark energy which will ensure that $R_h = t$ at all times, as at some point during the early expansion of the universe, 
an unphysical value of $\omegaw$ is required.

While beyond the scope of this paper, one could envisage a model with no matter in the early part of the universe, with a purely 
$\omegaw = -\frac{1}{3}$ dark energy component, which evolves at later times into a fraction of matter and a remaining dark energy 
component that maintains the required expansion form. While this would have significant impact on nucleosynthesis and structure 
formation, it is also extremely contrived and is quite removed from one of the claimed benefits of the $R_h = t$ universe, namely its
supposed simplicity in comparison to \cdm~\citep{2012arXiv1205.2713M}.

\section{Conclusions}\label{conclusions}
In this paper, we have examined the properties of the proposed $R_h = t$ universe, considering the evolution 
for the Cosmic Horizon in the presence of matter. It is found that for any density of matter, the resulting evolution 
deviated from the strictly required form demanded by the $R_h = t$ universe. This is imply a manifestation 
of universes becoming matter dominated in their earliest epochs. 

When considering the observational consequences of the presence of matter in a $R_h = t$ universe, it is
clear that this deviation significantly influence the key properties touted as being evidence for the 
efficacy of this cosmology, showing claims about the simplicity of the redshift-lookback time and
the removal of the need for inflation are, in fact, incorrect. 

Finally, we considered whether allowing an evolving equation of state for the dark energy component can
save the $R_h = t$ universe, finding that the presence of any matter results in the need for of an unphysical
equation of state in the early universe. While we could imagine some contrived models, where the equation of 
state of normal matter is forced to evolve, we conclude that the presence of matter in the universe is a blow 
to the proposed $R_h = t$ cosmology.

\section*{Acknowledgments}
The author thanks Luke Barnes and Krzysztof Bolejko  for interesting discussions. 
The anonymous referee is thanked for their  comments which improved this paper.

% References


\begin{thebibliography}{99}
%%
\bibitem[\protect\citeauthoryear{Bikwa, Melia, 
\& Shevchuk}{2012}]{2012MNRAS.421.3356B} Bikwa O., Melia F., Shevchuk A., 2012, MNRAS, 421, 3356
%%
\bibitem[\protect\citeauthoryear{Bilicki 
\& Seikel}{2012}]{2012MNRAS.425.1664B} Bilicki M., Seikel M., 2012, MNRAS, 425, 1664 
%%
\bibitem[\protect\citeauthoryear{Davis 
\& Lineweaver}{2004}]{2004PASA...21...97D} Davis T.~M., Lineweaver C.~H., 2004, PASA, 21, 97
%%
\bibitem[\protect\citeauthoryear{Ellis 
\& Rothman}{1993}]{1993AmJPh..61..883E} Ellis G.~F.~R., Rothman T., 1993, AmJPh, 61, 883 
%%
\bibitem[\protect\citeauthoryear{Harrison}{1991}]{1991ApJ...383...60H} 
Harrison E., 1991, ApJ, 383, 60
%%
%\bibitem[\protect\citeauthoryear{Harrison}{1993}]{1993ApJ...406..383H} 
%Harrison E., 1993, ApJ, 406, 383
%%
\bibitem[\protect\citeauthoryear{Lewis 
\& van Oirschot}{2012}]{2012MNRAS.423L..26L} Lewis G.~F., van Oirschot P., 2012, MNRAS, 423, L26 
%%
\bibitem[\protect\citeauthoryear{Lewis}{2013}]{2013arXiv1301.0305L} Lewis 
G.~F., 2013, arXiv, arXiv:1301.0305 
%%
\bibitem[\protect\citeauthoryear{Lima}{2007}]{2007arXiv0708.3414L} Lima 
J.~A.~S., 2007, arXiv, arXiv:0708.3414
%%
\bibitem[\protect\citeauthoryear{Linder}{1988}]{1988A&A...206..175L} Linder E.~V., 1988, A\&A, 206, 175 
%%
\bibitem[\protect\citeauthoryear{Melia}{2007}]{2007MNRAS.382.1917M} Melia 
F., 2007, MNRAS, 382, 1917 
%%
\bibitem[\protect\citeauthoryear{Melia}{2009}]{2009IJMPD..18.1113M} Melia 
F., 2009, IJMPD, 18, 1113 
%%
\bibitem[\protect\citeauthoryear{Melia 
\& Abdelqader}{2009}]{2009IJMPD..18.1889M} Melia F., Abdelqader M., 2009, IJMPD, 18, 1889 
%%
\bibitem[\protect\citeauthoryear{Melia 
\& Shevchuk}{2012}]{2012MNRAS.419.2579M} Melia F., Shevchuk A.~S.~H., 2012, MNRAS, 419, 2579 
%%
\bibitem[\protect\citeauthoryear{Melia}{2012a}]{2012JCAP...09..029M} Melia 
F., 2012a, JCAP, 9, 29 
%%
\bibitem[\protect\citeauthoryear{Melia}{2012b}]{2012AJ....144..110M} Melia 
F., 2012b, AJ, 144, 110 
%%
\bibitem[\protect\citeauthoryear{Melia}{2012c}]{2012arXiv1206.6527M} Melia 
F., 2012c, arXiv, arXiv:1206.6527 
%%
\bibitem[\protect\citeauthoryear{Melia}{2012d}]{2012arXiv1205.2713M} Melia 
F., 2012d, Australian Physics (May 2012), 83, arXiv, arXiv:1205.2713
%%
\bibitem[\protect\citeauthoryear{Melia}{2012e}]{2012arXiv1207.0734M} Melia 
F., 2012e, arXiv, arXiv:1207.0734 
%%
\bibitem[\protect\citeauthoryear{Melia}{2012f}]{2012arXiv1207.0015M} Melia 
F., 2012f, arXiv, arXiv:1207.0015 
%%
\bibitem[\protect\citeauthoryear{Melia}{2013}]{2013arXiv1301.0017M} Melia 
F., 2013, Accepted for publication in ApJ, arXiv, arXiv:1301.0017 
%
\bibitem[\protect\citeauthoryear{Mortlock et 
al.}{2011}]{2011Natur.474..616M} Mortlock D.~J., et al., 2011, Natur, 474, 
616 
\bibitem[Rindler(1956)]{1956MNRAS.116..662R} Rindler, W.\ 1956, MNRAS, 
116, 662 
%%
\bibitem[\protect\citeauthoryear{van Oirschot, Kwan, 
\& Lewis}{2010}]{2010MNRAS.404.1633V} van Oirschot P., Kwan J., Lewis G.~F., 2010, MNRAS, 404, 1633 
%%
\bibitem[\protect\citeauthoryear{Volonteri 
\& Rees}{2006}]{2006ApJ...650..669V} Volonteri M., Rees M.~J., 2006, ApJ, 650, 669
%%
\end{thebibliography}
\end{document}